\documentclass[pra,aps,showpacs,twocolumn]{revtex4-1}

\begin{document}
\title{Completely positive maps within the framework of direct-sum decomposition of state space}
\author{Longjiang Liu}
\affiliation{Department of Physics, Shandong University, Jinan 250100, People's Republic of China}
\author{D. M. Tong}
\email{tdm@sdu.edu.cn}
\affiliation{Department of Physics, Shandong University, Jinan 250100, People's Republic of China}
\date{\today}
\begin{abstract}
We investigate completely positive maps for an open system interacting with its environment. The families of the initial states for which the reduced dynamics can be described by a completely positive map are identified within the framework of direct-sum decomposition of state space. They includes not only separable states with vanishing or nonvanishing quantum discord but also entangled states. A general expression of the families as well as the Kraus operators for the completely positive maps are explicitly given. It significantly extends the previous results.
\end{abstract}
\maketitle

\section{introduction}
Any real quantum system $S$ inevitably interacts to some extent with its  environment $E$. The dynamics of an open system cannot be described by a unitary operator in general, although the combined system of the system and its environment experiences a unitary evolution, $\rho^{SE}(t)=U^{SE}(t)\rho^{SE}(0)U^{SE\dagger}(t)$, where $U^{SE}(t)$ denotes the unitary operator, $\rho^{SE}(0)$ and $\rho^{SE}(t)$ denote the initial state and the state at time $t$,
respectively. That is, a unitary operator does not exist to map the reduced state $\rho^S(0)\equiv \text{Tr}_E\rho^{SE}(0)$ to $\rho^S(t)\equiv \text{Tr}_E\rho^{SE}(t)$. To describe the dynamics of an open system, one used to assume that the combined system is initially in the product states,
\begin{eqnarray}
\rho^{SE}(0)=\rho^S\otimes\rho^E,\label{oldone}
\end{eqnarray}
where $\rho^E$ is a fixed density operator of the environment, and $\rho^S$ is an arbitrary density operator of the open system. In this case, the reduced dynamics of the open system can always be expressed as the Kraus representation, $\rho^S(t)=\sum_\mu K_\mu(t)\rho^S(0)K_{\mu}^{\dagger}(t)$, where $K_\mu(t)$ are called Kraus operators \cite{Kraus,Pechukas,Alicki,Nielsen,Preskill,Breuer}, dependent on the environment state $\rho^E$.  It is trace preserving if $\sum_\mu K_{\mu}^{\dagger}(t) K_\mu(t)=I$. A map from $\rho^{S}(0)$ to $\rho^{S}(t)$ is completely positive (CP) if and only if it can be expressed as the Kraus representation. Therefore, the reduced dynamics of the open system  is a CP map if $\rho^{SE}(0) \in\{\rho^S\otimes\rho^E\}$.

However, an open system may not be initially in a product state. Instead, many quantum systems of interest are initially correlated with its environment and the reduced dynamics with initial correlations has attracted much attentions \cite{Peter,Hayashi,Fonseca,Tong,Rodriguez-Rosario1,
Carteret,Shabani,Rodriguez-Rosario2,Zhang,Modi1,Devi,Tan,Masillo,Modi2,Xu,Brodutch,Smirne,Buscemi,Dominy}.
In the presence of initial correlations, the evolution of an open system may not have the Kraus representation that is valid to all the initial states, because an additional inhomogeneous part appears \cite{Peter}. To describe the reduced dynamics with the Kraus representation, it is necessary to specify the initial correlations of the system with its environment \cite{Peter,Hayashi,Fonseca,Tong}. There has been a growing interest in investigating the families of the initial states for which the reduced dynamics can be described by a CP map \cite{Rodriguez-Rosario1, Shabani,Brodutch,Buscemi,Dominy}. Particularly, it was recently proved that if the initial states are with the structure
\begin{eqnarray}
\rho^{SE}(0)=\sum_\alpha p_\alpha |\chi_\alpha^S\rangle \langle \chi_\alpha^S|
\otimes\rho_\alpha^E,\label{oldtwo}
\end{eqnarray}
where $\rho_\alpha^E$ is a fixed density operator on the environment space, $\{|\chi_\alpha^S\rangle \langle \chi_\alpha^S|\}$ is a fixed complete set of orthogonal projectors, and $p_\alpha$ is an arbitrary non-negative number satisfying $\sum_\alpha p_\alpha=1$, then the reduced dynamics can always be expressed as the Kraus
representation for the family of states $\{\sum_\alpha p_\alpha |\chi_\alpha^S\rangle \langle \chi_\alpha^S| \otimes\rho_\alpha^E\}$ with variable $p_\alpha$ \cite{Rodriguez-Rosario1}.  Therefore, the reduced dynamics of the open system  is a CP map if $\rho^{SE}(0) \in\{\sum_\alpha p_\alpha|\chi_\alpha^S\rangle \langle \chi_\alpha^S|\otimes\rho_\alpha^E\}$. Clearly, this work has relaxed the requirement of product states and extended the Kraus representation to a family of separable states with vanishing discord \cite{Ollivier,Henderson}.

Further extension was given in Ref. \cite{Shabani}, where it was stated that $\rho^S(0)\rightarrow \rho^S(t)$ is a CP map if the initial system-environment state $\rho^{SE}(0)$ has vanishing discord, i.e., can be written as
\begin{eqnarray}
\rho^{SE}(0)=\sum_{\alpha, k} \Pi_{\alpha k}^S \rho^{SE}\Pi_{\alpha k}^S ,\label{oldthree}
\end{eqnarray}
where $\{\Pi_{\alpha k}^S\}$ are one-dimensional projectors onto the eigenvectors of $\rho^S_\alpha$, and $\sum_k \Pi_{\alpha k}^S= \Pi_{\alpha}^S$ (see Ref. \cite{Shabani} for details).

It had been thought that vanishing quantum discord was necessary and sufficient condition for CP maps, but the authors of Ref. \cite{Brodutch} illustrated that vanishing quantum discord is not necessary for CP maps by presenting a counterexample, i.e., a family of separable states,
\begin{eqnarray}
\rho^{SE}(0)&=&\frac{p_1}{3}(|0\rangle\langle0|\otimes\rho_0^{E}+|1\rangle\langle1|\otimes\rho_1^{E}+
|+\rangle\langle+|\otimes\rho_+^{E})\nonumber\\
&&+\sum_{i=2}^{N_0}p_i|i\rangle\langle i|\otimes\rho_i^{E},\label{ce}
\end{eqnarray}
where $\rho_i^{E}~(i=0,1,\dots,N_0,+)$ are fixed density operators and $\sum_{i=1}^{N_0}p_i=1$. $\rho^{SE}(0)$ has nonvanishing discord except for the special cases of $p_1=0$ or $\rho_0^E=\rho_1^E=\rho_+^E$.  This implies that the initial states for which the reduced dynamics can be described by a completely positive map are beyond the family defined by Eqs. (\ref{oldtwo}) as well as (\ref{oldthree}).

The author of Ref. \cite{Buscemi} used a quantum date-processing inequality to investigate the existence of CP maps for initial system-environment correlations, and provided a complete characterization of the correlations that lead to CP reduced dynamics. It is shown that initial system-environment correlations always give rise to CP reduced dynamics if and only if a backward flow of information from the environment to the system cannot occur. By introducing a reference system $R$, choosing a fixed tripartite state $\rho^{RSE}$ with the conditional mutual information [$I(R:E|S)_{\rho}$] being zero, and steering states of $SE$ with linear maps on $R$,  a family of states $\rho^{SE}(0)$ can be obtained, where the reduced dynamics of the states in the family is CP for any system-environment interaction. More recently, the authors of Ref. \cite{Dominy} suggested a complete and consistent mathematical framework for the analysis of CP maps for correlated initial states. The problem of CP reduced dynamics is attracting broad interest.

It should be noted that the Kraus operators are necessarily dependent on the structure of initial states when the initial states are extended to the states with initial correlations. This is different from the case of the product states defined by Eq. (\ref{oldone}), where the Kraus operators are only dependent on the environment states but independent of the structure of system states. For example, the Kraus operators for the family defined by Eq. (\ref{oldtwo}) are dependent on $\rho^E_\alpha$ as well as $|\chi_\alpha^S\rangle \langle \chi_\alpha^S|$;
and the Kraus operators for the family defined by Eq. (\ref{ce}) are dependent on $\rho^E_0$, $\rho^E_1$, $\rho^E_+$, $\rho^E_i$, as well as $|0\rangle\langle 0|$, $|1\rangle\langle 1|$, $|+\rangle\langle +|$, $|i\rangle\langle i|$. However, such a CP map can sufficiently describe the reduced dynamics of the states in the family under consideration, although it is not applicable to the states outside the family. This is useful to the cases where the initial states of a quantum system are restricted to a certain family.

Completely positive maps play a fundamental role in the open quantum system's theory and are useful for quantum information processing \cite{Nielsen,Preskill}. Therefore, a full-out understanding on the initial correlations for which the reduced dynamics can be described by CP maps is instructive and necessary. In this paper, we offer an alternative perspective to describe the families of the initial states for which, for any system-environment interaction, there always exists a CP map describing the reduced system dynamics for all states in each family. We will show that those families can be described within the framework of direct-sum decomposition of state space. The merits of this framework are: $(1)$ all the previous results that are with an explicit expression, such as those given by Eqs. (\ref{oldone})-(\ref{ce}), can be put into the framework; $(2)$ more general families of the initial states for which the reduced dynamics is a CP map can be easily obtained by using the framework.

This paper is organized as follows. In Sec. \ref{cpsec}, we apply the framework of direct-sum decomposition of state space to the classification of initial states, and show that the previous results can be put into the framework. In Sec. \ref{cpsec2}, we extend the families of the initial states for which the reduced dynamics can be described by a CP map to a more general expression with the framework of direct-sum decomposition of state space. Section \ref{lastsec} is the conclusion.

\section{CP maps within the framework of direct-sum decomposition of space}\label{cpsec}
Before proceeding further, we first specify some notations used in this paper. $\mathcal{H}^S$ and $\mathcal{H}^E$ are used to represent the state space of an open system with $N$ dimensions and that of its environment with $M$ dimensions, respectively. $\{|\nu_i^S\rangle,~i=1,2,\dots,N\}$ and $\{|\mu_i^E\rangle, ~~i=1,2,\dots,M\}$ denote the basis of  $\mathcal{H}^S$ and $\mathcal{H}^E$, respectively. A direct-sum decomposition of the state space  $\mathcal{H}^S$ is denoted as $\mathcal{H}_1^S\oplus \mathcal{H}_2^S\oplus\cdots\oplus \mathcal{H}_{N_0}^S$, where the $\alpha-$th subspace $\mathcal{H}_\alpha^S$, $\alpha=1,2,\dots,N_0$, is  with $d_\alpha$ dimensions, and  $N_0$ is the total number of the subspaces in the decomposition. There is $\sum_{\alpha=1}^{N_0}d_\alpha=N$. $\Pi_\alpha^S$ is used to represent the projector of the $\alpha$-th subspace. They satisfy the relations $\Pi_\alpha^S\Pi_\beta^S
=\delta_{\alpha\beta}\Pi_\alpha^S$, and $\sum_{\alpha=1}^{N_0}\Pi_\alpha^S=I_N$. We further use $\rho_\alpha^S$, ~$\rho_\alpha^E$, and $\rho_\alpha^{SE}$ to denote the density operators defined on the $\alpha-$th subspace $\mathcal{H}_\alpha^S$, on the environment space $\mathcal{H}^E$, and on the direct product space $\mathcal{H}_\alpha^S\otimes \mathcal{H}^E$, respectively. For simplicity, we also use $\{|\nu_{\alpha i}^S\rangle, ~i=1,2,\dots,d_\alpha\}$ to denote the basis of the subspace $\mathcal{H}_\alpha^S$, which is a subset of $\{|\nu_i^S\rangle,~i=1,2,\dots,N\}$. There is then $\Pi_\alpha^S=\sum_{i=1}^{d_\alpha}|\nu_{\alpha
i}^S\protect\rangle\protect\langle \nu_{\alpha i}^S|$, and $\rho_\alpha^S$ can be expressed as $\rho_\alpha^S=\sum_{i,j=1}^{d_\alpha}\rho_{\alpha ij}^S|\nu_{\alpha i}^S\rangle\langle \nu_{\alpha j}^S|$, where $\rho_{\alpha ij}^S$ are parameters.

With these notations and relations, we start to extend the previous results. To make our result clear, we state it as two theorems.

\textit{Theorem 1.}  If the initial states of the combined system are with the structure,
\begin{eqnarray}
\rho^{SE}(0)=\sum_{\alpha=1}^{N_0}
p_\alpha\rho_\alpha^S\otimes\rho_\alpha^E,
\label{newone}
\end{eqnarray}
where $\rho_\alpha^E$ is a {\sl fixed} density operator on the environment space $\mathcal{H}^E$, $\rho_\alpha^S$ is an arbitrary density operator on the subspace $\mathcal{H}_\alpha^S$, and $p_\alpha$ is an arbitrary nonnegative number satisfying $\sum_{\alpha=1}^{N_0} p_\alpha=1$, then the reduced dynamics of the open system can be described by a CP map as long as $\rho^{SE}(0)\in \{\sum_{\alpha=1}^{N_0} p_\alpha\rho_\alpha^S\otimes\rho_\alpha^E\}$.

As mentioned above, CP maps are necessarily dependent on the structure of initial states when the states are with initial correlations. Therefore, the theorem are based on the framework of direct-sum decomposition of state space,
$\mathcal{H}_1^S\oplus \mathcal{H}_2^S\oplus\cdots\oplus \mathcal{H}_{N_0}^S $. It means that each subspace $H_\alpha^S$ is fixed, but $\rho_\alpha^S$ being with variable parameters $\rho_{\alpha ij}^S$  can take every density operators in the subspace. The reduced dynamics for all the states defined with the same decomposition $\mathcal{H}_1^S\oplus \mathcal{H}_2^S\oplus\cdots\oplus \mathcal{H}_{N_0}^S$ can be described by a CP map.

We now prove the theorem. To this end, we examine the relation between $\rho^S(t)= \text{Tr}_E\rho^{SE}(t)$ and
$\rho^S(0)=\text{Tr}_E\rho^{SE}(0)$. Note that $\rho_\alpha^E$ can be written as $\rho_\alpha^E=\sum_j\lambda_{\alpha j}|\phi_{\alpha j}^E\rangle\langle \phi_{\alpha j}^E|$, where $\lambda_{\alpha j}$ and $|\phi_{\alpha j}^E\rangle $ are the eigenvalues and eigenvectors of $\rho_\alpha^E$. By definition, we have
\begin{eqnarray}
\rho^S(t)&=&\textrm{Tr}_E\left[U^{SE}(t)\rho^{SE}(0)U^{SE\dag}(t)\right]\nonumber\\
&=&\sum_{\alpha i}p_\alpha
\langle\mu_i^E|U^{SE}\left(\rho_\alpha^S\otimes\rho_\alpha^E\right)U^{SE\dag}|\mu_i^E\rangle\nonumber\\
&=&\sum_{\alpha ij} \lambda_{\alpha
j}p_\alpha\langle\mu_i^E|U^{SE}\left(\rho_\alpha^S\otimes|\phi_{\alpha
j}^E\rangle\langle \phi_{\alpha j}^E|\right)U^{SE\dag}|\mu_i^E\rangle\nonumber\\
&=&\sum_{\alpha ij} \lambda_{\alpha j}p_\alpha\langle\mu_i^E|U^{SE}|\phi_{\alpha j}^E\rangle\rho_\alpha^S\langle\phi_{\alpha j}^E|U_{SE}^{\dag}|\mu_i^E\rangle,\nonumber\label{s1}
\end{eqnarray}
where $|\mu_i^E\rangle$, $i=1,2,\dots,M$, are a complete set of basis on $\mathcal{H}^E$. Here, we have used $U^{SE}$ to present $U^{SE}(t)$ for the sake of brevity.  Since the density operator $\rho_\alpha^S$ belongs to the subspace $\mathcal{H}_\alpha^S$ and $\Pi_\alpha^S$ is the projector on this subspace, we have $\Pi_\alpha^S\rho_\beta^S\Pi_\alpha^S=\delta_{\alpha\beta}\rho_\alpha^S$ and hence $\Pi_\alpha^S\rho^S(0)\Pi_\alpha^S=p_\alpha\rho_\alpha^S$, where $\rho^S(0)=\text{Tr}_E\rho^{SE}(0)=\sum_\alpha
p_\alpha\rho_\alpha^S$. By using these relations, we further get
\begin{eqnarray}
\rho^S(t)=\sum_{\alpha ij} \lambda_{\alpha j}\langle\mu_i^E|U^{SE}|\phi_{\alpha
j}^E\rangle\Pi_\alpha^S\rho^S(0)\Pi_\alpha^S\langle \phi_{\alpha
j}^E|U^{SE\dag}|\mu_i^E\rangle.\nonumber \label{s2}
\end{eqnarray}
Let
\begin{eqnarray}
K_{\alpha ij}(t)=\sqrt{\lambda_{\alpha j}}\langle\mu_i^E|U^{SE}|\phi_{\alpha j}^E\rangle\Pi_\alpha^S,  \label{m1}
\end{eqnarray}
with $\alpha=1,2,\dots,N_0$ and $i,j=1,2,\dots,M$. It is easy to verify that $\sum\limits_{\alpha ij}K_{\alpha ij}^\dag(t)K_{\alpha ij}(t)=I_N$. We finally obtain the Kraus representation,
\begin{eqnarray}
\rho^S(t)=\sum\limits_{\alpha ij}K_{\alpha ij}(t)\rho^S(0)K_{\alpha
ij}^\dagger(t). \label{k1}
\end{eqnarray}
Therefore,  the reduced dynamics of the open system  is a CP map for the family of initial states defined by Eq. (\ref{newone}). The above proof shows that $K_{\alpha ij}(t)$ are independent of $\rho_\alpha^S$ and $p_\alpha$, but they are dependent on $\Pi_\alpha^S$,  $\lambda_{\alpha j}$ and $|\phi_{\alpha j}^E\rangle$. $\Pi_\alpha^S$ are completely determined by the given framework $\mathcal{H}_1^S\oplus \mathcal{H}_2^S\oplus\cdots\oplus\mathcal{H}_{N_0}^S $.

It is worth noting that the previous results expressed by Eqs. (\ref{oldone}) and (\ref{oldtwo}) can be taken as special cases of \text{Theorem 1} with special direct-sum decompositions $\mathcal{H}_1^S\oplus \mathcal{H}_2^S\oplus\cdots\oplus\mathcal{H}_{N_0}^S$. Indeed, if we let $N_0=1$ and $d_1=N$, Eq. (\ref{newone}) becomes $\rho^{SE}(0)=\rho^S\otimes\rho^E$, which is just the well-known result Eq. (\ref{oldone}). If we let $N_0=N$ and $d_1=d_2=\cdots=d_N=1$, since $\rho_\alpha^S$ must be equal to $\Pi_\alpha^S$ in one-dimensional subspaces, Eq. (\ref{newone}) becomes $\rho^{SE}(0)= \sum_{\alpha=1}^{N}p_\alpha\Pi_\alpha^S \otimes \rho_\alpha^E$. It is just the result expressed by Eq. (\ref{oldtwo}) as  $\Pi_\alpha^S$ can be equivalently expressed as $|\chi_\alpha^S\rangle \langle \chi_\alpha^S|$ in the case of one-dimensional subspaces.  In general cases, if one or more $d_\alpha\neq 1$,  the family defined by $\rho^{SE}(0)=\sum_{\alpha=1}^{N_0} p_\alpha\rho_\alpha^S\otimes\rho_\alpha^E$ contains more states than that defined by $\rho^{SE}(0)=\sum_{\alpha=1}^N p_\alpha |\chi_{\alpha}^S\rangle \langle\chi_{\alpha}^S|\otimes\rho_\alpha^E$. This is because that $\rho_\alpha^S$ in the former family can be taken all the states in the subspace $\mathcal{H}_\alpha^S$ while its counterpart in the latter family is constrained by
$\rho_\alpha^S=\sum_{\beta=\beta_\alpha+1}^{\beta_\alpha+d_\alpha}c_{\beta} |\chi_{\beta}^S\rangle \langle \chi_{\beta}^S|$, where $0\leq c_\beta \leq 1$, satisfying $\sum_{\beta=\beta_\alpha+1}^{\beta_\alpha+d_\alpha}c_\beta=1$, and $\beta_\alpha=\sum_{i=1}^{\alpha-1}d_i$ with $\beta_1=0$. It is also worth noting that if we interpret the framework of direct-sum decomposition of state space in the notion of block diagonal matrix in Ref. \cite{Shabani} and match the $\alpha$-th subspace
$\mathcal{H}_\alpha^S$ with the $\alpha$-th block $\Phi^{\alpha}$, then \text{Theorem 1} can lead to the result expressed by Eq. (\ref{oldthree}).

We now take a concrete example as an illustration of \text{Theorem 1}. We consider a $4\times2$ combined system.
$|\nu^S\rangle$, $\nu=1,2,3,4$,  and $|\mu^E\rangle$, $\mu=1,2$, are used to denote the basis of $\mathcal{H}^S$ and $\mathcal{H}^E$, respectively. Two subspaces $\mathcal{H}_1^S$ and $\mathcal{H}_2^S$ are defined by $\Pi_1^S=|1^S\rangle\langle1^S|+|2^S\rangle\langle2^S|$ and $\Pi_2^S=|3^S\rangle\langle3^S|+|4^S\rangle\langle4^S|$, respectively. Let $\rho_1^E=|1^E\rangle\langle1^E|$ and $\rho_2^E=|2^E\rangle\langle 2^E|$. Then, the family of initial states defined by Eq. (\ref{newone}) is given as
\begin{eqnarray}
\rho^{SE}(0)=(1-p)\rho_1^{S}\otimes|1^E\rangle\langle1^E|+p\rho_2^S\otimes|2^E\rangle\langle2^E|,\label{t1}
\end{eqnarray}
where $0\leq p\leq1$, and $\rho_1^{S}$ and $\rho_2^{S}$ can take all the states in the subspaces $\mathcal{H}_1^S$ and $\mathcal{H}_2^S$, respectively. According to the theorem, the reduced dynamics for all the states expressed as Eq. (\ref{t1}) with variable $p$, $\rho_1^{S}$, and $\rho_2^{S}$ can be described by a CP map. The Kraus representation reads
\begin{eqnarray}
\rho^{S}(t)=\sum_{\alpha,i,j=1}^2K_{\alpha ij}(t)\rho^S(0)K_{\alpha ij}^\dagger(t),\label{example1}
\end{eqnarray}
with $\rho^S(0)=(1-p)\rho_1^{S}+p\rho_2^S$, where the nonzero Kraus operators are given by Eq. (\ref{m1}) (see Appendix \ref{a}).

\section{Extension of CP maps to general structure of initial states}\label{cpsec2}

\text{Theorem 1} concerns only the family of initial states with vanishing discord. All the density operators in each subspace $\{\mathcal{H}_\alpha^S\otimes\mathcal{H}_E \}$ in \text{Theorem 1}, i.e. the terms on the right side of Eq. (\ref{newone}), are with the product form $\rho_\alpha^S\otimes\rho_\alpha^E$. An interesting question is: Does there still exist a CP map if the density operators in some subspaces cannot be written as the product form? In other words, can the reduced dynamics for a state set including nonvanishing discord states or entangled states be described by a CP map? We find that it can be done if the non-product density operators are fixed. In this section, we extend the family of initial states to that including the separable states with nonvanishing discord as well as entangled states. We state this extension as \text{Theorem 2}.

\textit{Theorem 2.} If the initial states of the combined system are with the structure,
\begin{eqnarray}
\rho^{SE}(0)=\sum_{\alpha=1}^n
p_\alpha\rho_\alpha^{SE}+\sum_{\alpha=n+1}^{N_0}
p_\alpha\rho_\alpha^S\otimes\rho_\alpha^E,
\label{newtwo}
\end{eqnarray}
where $\rho_\alpha^{SE}$ is a {\sl fixed} density operator with nonvanishing discord on the direct product space $\mathcal{H}_\alpha^S\otimes \mathcal{H}^E$, $\rho_\alpha^E$ is a {\sl fixed} density operator on the environment space $\mathcal{H}^E$, $\rho_\alpha^S$ is an arbitrary density operator on the subspace $\mathcal{H}_\alpha^S$, and
$p_\alpha$ is an arbitrary non-negative number satisfying $\sum_{\alpha=1}^{N_0} p_\alpha=1$, then the reduced dynamics of the open system can be described by a CP map as long as $\rho^{SE}(0)\in \{\sum_{\alpha=1}^n p_\alpha\rho_\alpha^{SE}+\sum_{\alpha=n+1}^{N_0} p_\alpha\rho_\alpha^S\otimes\rho_\alpha^E\}$.

Clearly, \text{Theorem 2} is a generalization of \text{Theorem 1}. In expression (\ref{newtwo}), the number $n$ may take $0,1,\dots,N_0$, where the case of $n=0$ is just corresponding to \textit{Theorem 1}. To prove the theorem, we examine the relation between $\rho^S(t)$ and $\rho^S(0)$. From Eq. (\ref{newtwo}),  we have $\rho^S(0)= \text{Tr}_E\rho^{SE}(0)=\sum_{\alpha=1}^n p_\alpha\text{Tr}_E\rho_\alpha^{SE}+\sum_{\alpha=n+1}^{N_0} p_\alpha\rho_\alpha^S$. By definition, we have
\begin{eqnarray}
\rho^S(t)&=&\textrm{Tr}_E\left[U^{SE}(t)\rho^{SE}(0)U^{SE\dag}(t)\right]\nonumber\\
&=&\sum_{\alpha=1}^n \sum_{i=1}^M p_\alpha  \langle\mu_i^E|U^{SE}\rho_\alpha^{SE}U^{SE\dag}|\mu_i^E\rangle\nonumber\\
&&+\sum_{\alpha=n+1}^{N_0}\sum_{i=1}^M p_\alpha \langle\mu_i^E|U^{SE}\rho_\alpha^S\otimes\rho_\alpha^EU^{SE\dag}|\mu_i^E\rangle,\label{f1}
\end{eqnarray}
where $|\mu_i^E\rangle$, $i=1,2,\dots,M$, are a complete set of basis of $\mathcal{H}^E$.

To calculate the first term in the last line of Eq. (\ref{f1}), we use the spectral decomposition, $\rho_\alpha^{SE}=\sum_L\eta_{\alpha L}|\Psi_{\alpha L}^{SE}\rangle\langle\Psi_{\alpha L}^{SE}|$, where $\eta_{\alpha L}$ and $|\Psi_{\alpha L}^{SE}\rangle$ represent the eigenvalues and eigenvectors of $\rho_\alpha^{SE}$. Besides, since $\rho_\alpha^{SE}$ belongs to the space $\mathcal{H}_\alpha^S\otimes \mathcal{H}^E$ and therefore the reduced density operator $\text{Tr}_E\rho_\alpha^{SE}$ belongs to $\mathcal{H}_\alpha^S$, we then have $\sum_{i=1}^{d_\alpha}\langle \nu_{\alpha i}^S|\text{Tr}_E\rho_\beta^{SE}|\nu_{\alpha i}^S\rangle=\delta_{\alpha \beta}$, where $\{|\nu_{\alpha i}^S\rangle\}$ represents the basis of $\mathcal{H}_\alpha^S$ with $\sum_{i=1}^{d_\alpha}|\nu_{\alpha i}^S\rangle \langle\nu_{\alpha i}^S|=\Pi_\alpha^S$. It further leads to $\sum_{i=1}^{d_\alpha}\langle \nu_{\alpha i}^S|\rho^S(0)|\nu_{\alpha i}^S\rangle=\sum_{i=1}^{d_\alpha}\langle \nu_{\alpha i}^S|\text{Tr}_E\rho^{SE}(0)|\nu_{\alpha i}^S\rangle =p_\alpha$. By using these relations, we get
\begin{eqnarray}
&& p_\alpha  \langle\mu_i^E|U^{SE}\rho_\alpha^{SE}U^{SE\dag}|\mu_i^E\rangle\nonumber\\
&&=\sum_{L=1}^{d_\alpha \times
M}p_\alpha\eta_{\alpha L} \langle\mu_i^E|U^{SE}|\Psi_{\alpha L}^{SE}\rangle\langle\Psi_{\alpha L}^{SE}|U^{SE\dag}|\mu_i^E\rangle\nonumber\\
&&=\sum_{L=1}^{d_\alpha \times
M}\sum_{k=1}^{d_\alpha}\left[\eta_{\alpha L}
\langle\mu_i^E|U^{SE}|\Psi_{\alpha L}^{SE}\rangle\langle
\nu_{\alpha k}^S|\rho^S(0)|\nu_{\alpha k}^S\rangle\right.\nonumber\\
&&~~~~~~~~~~~~~~~~~\left.\times \langle\Psi_{\alpha
L}^{SE}|U^{SE\dag}|\mu_i^E\rangle\right].\label{f2}
\end{eqnarray}

For the second term in the last line of Eq. (\ref{f1}), by following the approach used in the proof of \text{Theorem 1}, we have
\begin{eqnarray}
&& p_\alpha
\langle\mu_i^E|U^{SE}\rho_\alpha^S\otimes\rho_\alpha^EU^{SE\dag}|\mu_i^E\rangle\nonumber\\
&&=\sum_{j=1}^M\lambda_{\alpha j}p_\alpha\langle\mu_i^E|U^{SE}\rho_\alpha^S\otimes|\phi_{\alpha j}^E\rangle\langle \phi_{\alpha j}^E|U^{SE\dag}|\mu_i^E\rangle\nonumber\\
&&=\sum_{j=1}^M \lambda_{\alpha j}p_\alpha\langle\mu_i^E|U^{SE}|\phi_{\alpha j}^E\rangle\rho_\alpha^S\langle \phi_{\alpha j}^E|U_{SE}^{\dag}|\mu_i^E\rangle\nonumber\\
&&=\sum_{j=1}^M\lambda_{\alpha j}\langle\mu_i^E|U^{SE}|\phi_{\alpha j}^E\rangle\Pi_\alpha^S\rho^S(0)\Pi_\alpha^S\langle \phi_{\alpha j}^E|U^{SE\dag}|\mu_i^E\rangle,\label{f3}
\end{eqnarray}
where $\lambda_{\alpha j}$ and $|\phi_{\alpha j}^E\rangle$ are the eigenvalues and eigenvectors of $\rho_\alpha^E$.

Substituting Eqs. (\ref{f2}) and  (\ref{f3}) into (\ref{f1}), we have
\begin{eqnarray}
\rho^S(t)&=&\sum_{\alpha=1}^n \sum_{i=1}^M \sum_{L=1}^{d_\alpha \times
M}\sum_{k=1}^{d_\alpha}\left[\eta_{\alpha L} \langle\mu_i^E|U^{SE}|\Psi_{\alpha L}^{SE}\rangle\langle \nu_{\alpha k}^S|\right.\nonumber\\
&&~~~~~~~~~~~~~~~~~~~~~~\times \rho^S(0)|\nu_{\alpha k}^S\rangle\langle\Psi_{\alpha L}^{SE}|U^{SE\dag}|\mu_i^E\rangle\left.\right]\nonumber\\
&&+\sum_{\alpha=n+1}^{N_0}\sum_{i,j=1}^M\left[\right.\lambda_{\alpha j}\langle\mu_i^E|U^{SE}|\phi_{\alpha
j}^E\rangle\Pi_\alpha^S\rho^S(0)\Pi_\alpha^S\nonumber\\
&&~~~~~~~~~~~~~~~~~~\times \langle \phi_{\alpha j}^E|U^{SE\dag}|\mu_i^E\rangle\left.\right] .\label{f4}
\end{eqnarray}
Let
\begin{eqnarray}
K_{\alpha iLk}(t)=\sqrt{\eta_{\alpha L}}\langle\mu_i^E|U^{SE}|\Psi_{\alpha L}^{SE}\rangle\langle \nu_{\alpha k}^S|, ~~\alpha=1,\dots,n, \label{k21}
\end{eqnarray}
and
\begin{eqnarray}
 K_{\alpha ij}(t)=\sqrt{\lambda_{\alpha j}}\langle\mu_i^E|U^{SE}|\phi_{\alpha j}^E\rangle\Pi_\alpha^S,~~\alpha=n+1,\dots,N_0, \label{k22}
\end{eqnarray}
with  $i,~j=1,2,\dots,M$, $k=1,2,\dots,d_\alpha$ and $L=1,2,\dots,d_\alpha\times M$. It is easy to verify that $\sum_{\alpha=1}^n \sum_{i,L,k} K_{\alpha iLk}^\dagger K_{\alpha iLk}+ \sum_{\alpha=n+1}^{N_0}\sum_{i,j} K_{\alpha ij}^\dagger K_{\alpha ij}= \sum_{\alpha=1}^{n} \sum_k|\nu_{\alpha k}^S\rangle\langle \nu_{\alpha k}^S| + \sum_{\alpha=n+1}^{N_0}\Pi_\alpha=  \sum_{\alpha=1}^{N_0}\Pi_\alpha= I_N$. We finally obtain the Kraus representation,
\begin{eqnarray}
\rho^S(t)&=&\sum_{\alpha=1}^n \sum_{i=1}^M \sum_{L=1}^{d_\alpha \times
M}\sum_{k=1}^{d_\alpha} K_{\alpha iLk}(t)\rho^S(0)K_{\alpha iLk}^\dagger(t)\nonumber\\
 &&+ \sum_{\alpha=n+1}^{N_0}\sum_{i,j=1}^M K_{\alpha ij}(t)\rho^S(0)K_{\alpha ij}^\dagger(t). \label{repre2}
\end{eqnarray}
Therefore,  the reduced dynamics of the open system  is a CP map if $\rho^{SE}(0) \in\{\sum_{\alpha=1}^n
p_\alpha\rho_\alpha^{SE}+\sum_{\alpha=n+1}^{N_0} p_\alpha\rho_\alpha^S\otimes\rho_\alpha^E\}$, as defined by Eq.
(\ref{newtwo}). We should stress that the density operators $\rho_\alpha^{SE}$ and  $\rho_\alpha^E$ in the theorem must be fixed. $\rho_\alpha^S$ can take all the density operators in the subspace $\mathcal{H}_\alpha^S$, and $p_\alpha$ is an arbitrary non-negative number satisfying $\sum_{\alpha=1}^{N_0} p_\alpha=1$. The reduced dynamics for all the initial states expressed by Eq. (\ref{newtwo}) with variable $\rho_\alpha^S$ and $p_\alpha$ but fixed $\rho_\alpha^{SE}$ and $\rho_\alpha^E$ can be described by a CP map defined by Eqs. (\ref{k21})-(\ref{repre2}). Besides, we would like to point out that the dependence of $K_{\alpha iLk}(t)$ on $|\nu_{\alpha k}^S\rangle$ does not weaken the validity of the CP map. $\{|\nu_{\alpha k}^S\rangle,~k=1,2,\dots,d_\alpha\}$ is the basis of the subspace $\mathcal{H}_\alpha^S$. Although the basis of $\mathcal{H}_\alpha^S$ is not unique, Kraus operators obtained by using a different basis are equivalent up to a unitary transformation, and therefore different choices of the basis give the same CP map.

It is worth noting that the result expressed by Eq. (\ref{ce}) can be taken as a special case of \text{Theorem 2}. If we define $\mathcal{H}_1^S$ and $\mathcal{H}_\alpha^S$ by $\Pi_1^S=|0\rangle\langle0|+|1\rangle\langle1|$ and $\Pi_\alpha^S=|\alpha\rangle\langle \alpha|$, $\alpha=2,3,\dots,N_0$, and let $\rho_1^{SE}=(|0\rangle\langle0|\otimes\rho_0^{E}+|1\rangle\langle1|\otimes\rho_1^{E}+
|+\rangle\langle+|\otimes\rho_+^{E})/3$ and $\rho_{\alpha}^S=|\alpha\rangle\langle \alpha|$, then Eq. (\ref{newtwo}) becomes Eq. (\ref{ce}). Therefore, all the previous results described by Eqs. (\ref{oldone})-(\ref{ce}) can be taken as special cases of \text{Theorem 2}, as the results described by Eqs. (\ref{oldone})-(\ref{oldthree}) can be obtained from \text{Theorem 1}, which is a special case of \text{Theorem 2} at $n=0$. Besides, since the reduced dynamics of product states can always be described as a CP map, the term $\sum_{\alpha=1}^n
p_\alpha\rho_\alpha^{SE}$ in Eq. (\ref{newtwo}) of \text{Theorem 2} can be recast as a more general form if $\mathcal{H}^S_\alpha =\mathcal{H}^S_{\alpha 1} \otimes \mathcal{H}^S_{\alpha 2}$, i.e., if the $\alpha-$th subspace can be written as a tensor product of two smaller subspaces. Indeed, it is easy to verify that the theorem is still valid if Eq. (\ref{newtwo}) is replaced by $\rho^{SE}(0)=\sum_{\alpha=1}^n p_\alpha \rho^S_{\alpha 1} \otimes \rho^{SE}_{\alpha 2} + \sum_{\alpha=n+1}^{N_0} p_\alpha\rho_\alpha^S \otimes \rho_\alpha^E$, where $\rho^S_{\alpha 1}$ is an arbitrary density operator on the subspace $\mathcal{H}^S_{\alpha 1}$ and $\rho^{SE}_{\alpha 2}$ is a {\sl fixed} density operator with nonvanishing discord on
the space $\mathcal{H}^S_{\alpha 2}\otimes \mathcal{H}^E$.

Considering that all the families defined by Eqs. (\ref{oldone})-(\ref{ce}) do not include any entangled state,  we would like to present two more  examples, which are related to entangled states. First, if we let $n=N_0$, Eq. (\ref{newtwo}) reduces to $\rho^{SE}(0)=\sum_{\alpha=1}^{N_0} p_\alpha\rho_\alpha^{SE}$, and all the states in this family are entangled as long as each $\rho_\alpha^{SE}$ is taken to be entangled. As an example, we continue to consider a $4\times2$ combined system. The two subspaces $\mathcal{H}_1$ and $\mathcal{H}_2$ are still defined by $\Pi_1^S=|1^S\rangle\langle1^S|+|2^S\rangle\langle2^S|$ and $\Pi_2^S=|3^S\rangle\langle3^S|+|4^S\rangle\langle4^S|$, respectively. Let $\rho_1^{SE}=|\Psi_{11}^{SE}\rangle\langle\Psi_{11}^{SE}|$ and  $\rho_2^{SE}=|\Psi_{21}^{SE}
\rangle\langle\Psi_{21}^{SE}|$, where $|\Psi_{11}^{SE}\rangle=(|1^S1^E\rangle+ |2^S2^E\rangle)/\sqrt{2}$ and $|\Psi_{21}^{SE}\rangle=(|3^S1^E\rangle+ |4^S2^E\rangle)/\sqrt{2}$. Then, the initial states defined by Eq. (\ref{newtwo}) are given as
\begin{eqnarray}
\rho^{SE}(0)=p|\Psi_{11}^{SE}\rangle\langle\Psi_{11}^{SE}|+(1-p)|\Psi_{21}^{SE}
\rangle\langle\Psi_{21}^{SE}|, \label{t2}
\end{eqnarray}
where $0\leq p\leq 1$.  The reduced dynamics for all the states with variable $p$ can be described by a CP map. The Kraus representation reads
\begin{eqnarray}
\rho^{S}(t)=\sum_{\alpha,i,k=1}^2 \sum_{L=1}^{4}K_{\alpha iLk}(t)\rho^S(0)K_{\alpha iLk}^\dagger(t),\label{example2}
\end{eqnarray}
with $\rho^{S}(0)=\left[p|1^S\rangle\langle1^S|+p|2^S\rangle\langle2^S|+(1-p)|3^S\rangle\langle3^S|+\right.$\\
$\left.(1-p)|4^S\rangle\langle4^S|\right]/2$, where the nonzero Kraus operators are given by Eq. (\ref{k21}) (see Appendix \ref{b}). Second, we consider a general case with $1\leq n \leq N_0-1$. In this case, the family of the initial states for which the reduced dynamics can be described by a CP map may consist of separable states with vanishing discord, separable states with nonvanishing discord, and entangled states. For example, we consider
a $6\times2$ combined system. $|\nu^S\rangle$, $\nu=1,2,\dots,6$, and $|\mu^E\rangle$, $\mu=1,2$, are used to denote the basis of $\mathcal{H}^S$ and $\mathcal{H}^E$, respectively. The three subspaces $\mathcal{H}_1^S$, $\mathcal{H}_2^S$, and $\mathcal{H}_3^S$ are defined by $\Pi_1^S=|1^S\rangle\langle1^S|+|2^S\rangle\langle2^S|$,
$\Pi_2^S=|3^S\rangle\langle3^S|+|4^S\rangle\langle4^S|$, and $\Pi_3^S=|5^S\rangle\langle5^S|+|6^S\rangle\langle6^S|$,
respectively. Let $\rho_1^{SE}=(|1^S1^E\rangle\langle1^S1^E|+|+^S+^E\rangle\langle+^S+^E|)/2$,
$\rho_2^{SE}=(|3^S1^E\rangle+|4^S2^E\rangle)(\langle3^S1^E|+\langle4^S2^E|)/2$, and $\rho_3^{E}=|2^E\rangle\langle2^E|$, where $|+\rangle=(|1\rangle+|2\rangle)/\sqrt{2}$. Then, the initial states
defined by Eq. (\ref{newtwo}) are given as
\begin{eqnarray}
\rho^{SE}(0)=\sum_{i=1}^2p_i\rho_i^{SE}+p_3\rho_3^S\otimes|2^E\rangle\langle2^E|,
\end{eqnarray}
where $p_1+p_2+p_3=1$, and $\rho_3^S$ is an arbitrary state in the subspace $\mathcal{H}_3^S$.  The reduced dynamics for all the states with variable $p_1$, $p_2$, $p_3$, and $\rho_3^S$ can be described by a CP map. The Kraus representation reads
\begin{eqnarray}
\rho^{S}(t)&=&\sum_{\alpha,i,k=1}^2 \sum_{L=1}^{4} K_{\alpha iLk}(t)\rho^S(0)K_{\alpha iLk}^\dagger(t)\nonumber\\
 &&+ \sum_{i,j=1}^2 K_{3ij}(t)\rho^S(0)K_{3ij}^\dagger(t),\label{example3}
\end{eqnarray}
with $\rho^S(0)=p_1(|1^S\rangle\langle1^S|+|+^S\rangle\langle+^S|)/2+p_2(|3^S\rangle\langle3^S|+|4^S\rangle\langle4^S|)/2+p_3\rho_3^S$,
where the nonzero Kraus operators are given by Eqs. (\ref{k21}) and (\ref{k22}) (see Appendix \ref{c}).

It is also interesting to compare our result with that given in Ref.\cite{Buscemi}. The author of that paper
provides a general condition of the initial system-environment correlations for which there always exists a corresponding CP map describing the reduced system dynamics, while we give an explicit expression of the initial states for which the reduced dynamics can be described by a CP map. The condition given in that paper is based on the tripartite system consisting of the open system, its environment, and the reference system, while our discussion is based on the bipartite system consisting of the open system and its environment only. If a reference system adds to our bipartite system, the initial sates defined in our theorems will fulfill the condition in that paper. In this sense, our result is consistent with the physical insight first put forward in Ref.\cite{Buscemi}, and the families of the initial states described in the bipartite scenario can be derived from that in the tripartite scenario.

\section{Conclusion}\label{lastsec}

In conclusion, we put forward an alternative perspective to investigate CP maps of an open system interacting with its environment. The structure of the initial states are described  within the framework of direct-sum decomposition of state space. Within the framework, we have identified the families of the initial states for which the reduced dynamics can always be described by a CP map, regardless of what the unitary operators $U^{SE}$(t) are. Our main finding is described by $\text{Theorem 1}$ and $\text{Theorem 2}$, while \text{Theorem 1} can be regarded as a
special case of $\text{Theorem 2}$. It shows that the reduced dynamics of the open system is CP if the combined system is initially in the family of states defined by Eq. (\ref{newtwo}). All the previous results described by Eqs. (\ref{oldone})-(\ref{ce}) can be taken as special cases of \text{Theorem 2}.  We have further illustrated the theorem with some examples: all the initial states in a family being with vanishing discord, all the initial states in a family being with nonvanishing discord, and the initial states in a family consisting of separable states with vanishing discord, separable states with nonvanishing discord, and entangled states. Our work significantly extends the previous results.

\begin{acknowledgments}
This work was supported by NSF China with Grant No.11175105 and the Taishan Scholarship Project of Shandong Province.
\end{acknowledgments}

\appendix
\section{The nonzero Kraus operators of example one}\label{a}
The nonzero Kraus operators in Eq. (\ref{example1}) are
\begin{eqnarray}
&&K_{111}(t)=\langle1^E|U^{SE}|1^E\rangle\Pi_{1}^S,\nonumber\\
&&K_{121}(t)=\langle2^E|U^{SE}|1^E\rangle\Pi_{1}^S,\nonumber\\
&&K_{211}(t)=\langle1^E|U^{SE}|2^E\rangle\Pi_{2}^S,\nonumber\\
&&K_{221}(t)=\langle2^E|U^{SE}|2^E\rangle\Pi_{2}^S.\label{ak1}
\end{eqnarray}
\section{The nonzero Kraus operators of example two}\label{b}
The nonzero Kraus operators in Eq. (\ref{example2}) are
\begin{eqnarray}
&&K_{1111}(t)=\langle1^E|U^{SE}|\Psi_{11}^{SE}\rangle\langle1^S|,\nonumber\\
&&K_{1112}(t)=\langle1^E|U^{SE}|\Psi_{11}^{SE}\rangle\langle2^S|,\nonumber\\
&&K_{1211}(t)=\langle2^E|U^{SE}|\Psi_{11}^{SE}\rangle\langle1^S|,\nonumber\\
&&K_{1212}(t)=\langle2^E|U^{SE}|\Psi_{11}^{SE}\rangle\langle2^S|,\nonumber\\
&&K_{2111}(t)=\langle1^E|U^{SE}|\Psi_{21}^{SE}\rangle\langle3^S|,\nonumber\\
&&K_{2112}(t)=\langle1^E|U^{SE}|\Psi_{21}^{SE}\rangle\langle4^S|,\nonumber\\
&&K_{2211}(t)=\langle2^E|U^{SE}|\Psi_{21}^{SE}\rangle\langle3^S|,\nonumber\\
&&K_{2212}(t)=\langle2^E|U^{SE}|\Psi_{21}^{SE}\rangle\langle4^S|.\label{ak2}
\end{eqnarray}

\section{The nonzero Kraus operators of example three}\label{c}
The nonzero Kraus operators in Eq. (\ref{example3}) are
\begin{eqnarray}
&&K_{1111}(t)=\frac{\sqrt{3}}{2}\langle1^E|U^{SE}|\Psi_{11}^{SE}\rangle\langle 1^S|,\nonumber\\ &&K_{1112}(t)=\frac{\sqrt{3}}{2}\langle1^E|U^{SE}|\Psi_{11}^{SE}\rangle\langle2^S|,\nonumber\\ &&K_{1211}(t)=\frac{\sqrt{3}}{2}\langle2^E|U^{SE}|\Psi_{11}^{SE}\rangle\langle1^S|,\nonumber\\ &&K_{1212}(t)=\frac{\sqrt{3}}{2}\langle2^E|U^{SE}|\Psi_{11}^{SE}\rangle\langle2^S|,\nonumber\\ &&K_{1121}(t)=\frac{1}{2}\langle1^E|U^{SE}|\Psi_{12}^{SE}\rangle\langle1^S|,\nonumber\\ &&K_{1122}(t)=\frac{1}{2}\langle1^E|U^{SE}|\Psi_{12}^{SE}\rangle\langle2^S|,\nonumber\\ &&K_{1221}(t)=\frac{1}{2}\langle2^E|U^{SE}|\Psi_{12}^{SE}\rangle\langle1^S|,\nonumber\\ &&K_{1222}(t)=\frac{1}{2}\langle2^E|U^{SE}|\Psi_{12}^{SE}\rangle\langle2^S|,\nonumber\\ &&K_{2111}(t)=\langle1^E|U^{SE}|\Psi_{21}^{SE}\rangle\langle3^S|,\nonumber\\
&&K_{2112}(t)=\langle1^E|U^{SE}|\Psi_{21}^{SE}\rangle\langle4^S|,\nonumber\\
&&K_{2211}(t)=\langle2^E|U^{SE}|\Psi_{21}^{SE}\rangle\langle3^S|,\nonumber\\
&&K_{2212}(t)=\langle2^E|U^{SE}|\Psi_{21}^{SE}\rangle\langle4^S|,\nonumber\\
&&K_{311}(t)=\langle1^E|U^{SE}|2^E\rangle\Pi_3^S, \nonumber\\
&&K_{321}(t)=\langle2^E|U^{SE}|2^E\rangle\Pi_3^S.
\end{eqnarray}
Here,
$|\Psi_{11}^{SE}\rangle=(3|1^S1^E\rangle+|1^S2^E\rangle+|2^S1^E\rangle+|2^S2^E\rangle)/2\sqrt{3}$,
$|\Psi_{12}^{SE}\rangle=(-|1^S1^E\rangle+|1^S2^E\rangle+|2^S1^E\rangle+|2^S2^E\rangle)/2$, and $|\Psi_{21}^{SE}\rangle=(|3^S1^E\rangle+|4^S2^E\rangle)/\sqrt{2}$, which are obtained from the spectral decompositions,   $\rho_1^{SE}=\frac{3}{4}|\Psi_{11}^{SE}\rangle\langle\Psi_{11}^{SE}|
+\frac{1}{4}|\Psi_{12}^{SE}\rangle\langle\Psi_{12}^{SE}|$, and $\rho_2^{SE}=|\Psi_{21}^{SE}\rangle\langle\Psi_{21}^{SE}|$.

\end{document}